\documentclass[preprintnumbers, floatfix,letterpaper,aps,prd,epsfig,nofootinbib,
twocolumn
]{revtex4-1}
\usepackage{bm,graphicx,dcolumn,epstopdf,epsf, latexsym, mathbbol, amssymb,amsmath,color,slashed, mathrsfs,mathcomp, simplewick, natbib}
\usepackage[center]{subfigure}
\usepackage{multirow}
\usepackage{makecell}
\usepackage[colorlinks,linkcolor=blue,citecolor=blue,urlcolor=blue]{hyperref}

\begin{document}
	\allowdisplaybreaks
	\newcommand{\bq}{\begin{equation}}
	\newcommand{\eq}{\end{equation}}
	\newcommand{\bqn}{\begin{eqnarray}}
	\newcommand{\eqn}{\end{eqnarray}}
	\newcommand{\nb}{\nonumber}
	\newcommand{\lb}{\label}
	\newcommand{\f}{\frac}
	\newcommand{\p}{\partial}
	\newcommand{\PRL}{Phys. Rev. Lett.}
	\newcommand{\PLB}{Phys. Lett. B}
	\newcommand{\PRD}{Phys. Rev. D}
	\newcommand{\CQG}{Class. Quantum Grav.}
	\newcommand{\JCAP}{J. Cosmol. Astropart. Phys.}
	\newcommand{\JHEP}{J. High. Energy. Phys.}
	\newcommand{\red}{\textcolor{black}}
	
	\title{Observational tests of the self-dual spacetime in loop quantum gravity}
	
	
	\author{Tao Zhu${}^{a, b}$}
	\email{zhut05@zjut.edu.cn; Corresponding author}
	

	\author{Anzhong Wang${}^{c}$}
	\email{anzhong_wang@baylor.edu}
	
	\affiliation{${}^{a}$Institute for Theoretical Physics \& Cosmology, Zhejiang University of Technology, Hangzhou, 310023, China\\
		${}^{b}$ United Center for Gravitational Wave Physics (UCGWP),  Zhejiang University of Technology, Hangzhou, 310023, China\\
		${}^{c}$ GCAP-CASPER, Physics Department, Baylor University, Waco, Texas 76798-7316, USA}

	\date{\today}
	
\begin{abstract}

The self-dual spacetime was derived from the mini-superspace approach, based on the polymerization quantization procedure in loop quantum gravity (LQG). Its deviation from the Schwarzschild spacetime is characterized by the polymeric function $P$, purely due to the geometric quantum effects from LQG.  In this paper, we consider the observational constraints imposed on  $P$ by using the solar system experiments and observations. For this purpose, we calculate in detail the effects of   $P$ on astronomical observations conducted in the Solar system, including the deflection angle of light by the Sun, gravitational time delay, perihelion advance, and geodetic procession. The observational constraints are derived by confronting the theoretical predictions with the most recent observations. Among these constraints, we find that the tightest one comes from the measurement of the gravitational time delay by the Cassini mission, which yields $0<P<5.5\times 10^{-6}$. In addition, we also discuss the potential constraint that can be obtained   in the near future by the joint European-Japanese BepiColombo project, and show that it  could significantly improve the current constraints.

\end{abstract}

\maketitle
	
\section{Introduction}
\renewcommand{\theequation}{1.\arabic{equation}} \setcounter{equation}{0}

Einstein's theory of general relativity (GR) was proposed over a century ago and has successfully passed all the observational tests carried out so far. In the weak  field regime, GR was tested with various ground- and space-based precision experiments, including the three classical tests, namely the perihelion advance of Mercury, the deflection of light by the Sun, and the gravitational redshift \cite{will2014}. In the strong field regime, GR was also confronted with  observations of binary pulsar systems \cite{pulsar1, pulsar2}, the extraordinary observation of the M87* black hole shadow by the Event Horizon Collaboration \cite{EHT}, and observations of gravitational waves generated due to the merging of black holes 
and/or neutron stars by the LIGO experiment \cite{ligo}. These observations are all remarkably consistent with the predictions of GR. 

Despite of all these  successes, there are also various  reasons to believe that  GR may not be the complete theory of gravity. First, the accelerated expansion of the universe \cite{dark_energy1, dark_energy2, dark_energy3, dark_energy4} and the inconsistencies in galaxy rotation curves \cite{dark_energy4, dark_matter1,dark_matter2,dark_matter3,dark_matter4} are difficult to explain within the framework of GR without introducing dark energy and dark matter. Second, the standard inflationary paradigm in the early universe also suffers from the trans-Planckian problem \cite{inflation1, inflation2}. Third, Einstein's GR does not employ any quantum principles and it is still an unsolved question of unifying GR and quantum mechanics \cite{QG1,QG2}. In addition, GR inevitably leads to singularities both at the initial of the universe   \cite{singularity1, singularity2} and in the interiors of black hole spacetimes \cite{hawking}, at which our known physics laws become all invalid. All these issues indicate that  the classical GR might need to be modified.  In particular, the spacetime singularities ought to be resolved after quantum  gravitational effects are taken into account.

Recently, in the context of LQG, a spherical symmetric spacetime,  known as the LQG corrected Schwarzschild spacetime or self-dual spacetime, was constructed \cite{LQG_BH} \footnote{{In the last couple of years, loop quantum black holes (LQBHs) have been extensively studied, see, for instance, \cite{AOS18a,AOS18b,BBy18,ABP19,ADL20}.  For more details, we refer readers to the review articles, \cite{Perez17,Rovelli18,BMM18,Ashtekar20}.}}. In particular, it has been shown that this self-dual spacetime is regular and free of any spacetime curvature singularity. In the construction of the solution,  the minimum area of the full LQG is the fundamental ingredient to solve the black hole space-time singularity problem. Moreover,  the deviation of the self-dual spacetime from the Schwarzschild one can be  characterized by the minimal area and the Barbero-Immirzi parameter arising from LQG. As mentioned in \cite{Sahu:2015dea, Modesto:2009ve}, another important aspect of this solution is that it is self-dual in sense of T-duality. One can verify that under the transformation \red{ $r \to a_0/r$ with $a_0$ being related to the minimal area gap $A_{\rm min}$ of LQG via $a_0=A_{\rm min}/8\pi$},, the metric remains invariant, with suitable re-parameterization of other variables, hence marking itself as 
satisfying  the T-duality.

 An important question now is whether the LQG effects of the self-dual spacetime can leave any observational signatures for the current and/or forthcoming experiments, so LQG can be tested or constrained directly by observations. Such considerations have attracted a great deal of attention lately and several phenomenological implications of the self-dual spacetime have been already investigated \cite{Alesci:2011wn, Chen:2011zzi, Dasgupta:2012nk, Hossenfelder:2012tc, Barrau:2014yka, Sahu:2015dea, Cruz:2015bcj, add1, add2}. In particular,  the LQG effects on the shadow of the rotating black hole has been discussed in details and their observational implications to the latest Event Horizon Telescope (EHT) observation of the supermassive black hole, M87*,  has also been explored \cite{Liu:2020ola}. In addition, with the calculation of the gravitational lensing in the self-dual spacetime,  the polymeric function  has been constrained by using the Geodetic Very-Long-Baseline Interferometry Data of the solar gravitational deflection of Radio Waves \cite{Sahu:2015dea},  \red{which leads to a constraint on the polymeric parameter $\delta$ of LQG, $\delta < 0.1$}.

In this paper, we study the effects of LQG to observations conducted in the Solar System. We calculate in details the effects of the polymeric function in the self-dual spacetime to the light deflection by the Sun, the gravitational time delay,  perihelion advance, and geodetic procession for a spinning object. With these theoretical calculations, we derive the observational constraints from some recent observational datasets, including the VLBI observation of quasars, Cassini experiment, MESSENGER mission, LAGEOS satelite, observations of S2 star at Galactic center, Gravity Probe B, and the lunar laser ranging data. Among these constraints, we find that the tightest constraint comes from the measurement of the gravitational time delay by the Cassini mission. In addition, we also discuss the potential constraint that can be derived in the near future by the joint European-Japanese BepiColombo project. While more detections and experiments are continuously being carried out, it is expected that the constraints on the LQG effects  will be improved dramatically and deeper understanding of LQG will be achieved.  \red{At last, we would like to mention that we only consider the static self-dual spacetime in this paper and ignore the effects of the angular momentum of the spacetime. For all the observational effects we considered in this paper, the effects due to rotation of the Sun or Earth are expected to be very small.} 
 
The plan of the rest of the paper is as follows. In Sec. II, we present a very brief introduction to the self-dual spacetime, while in Sec. III, we first consider the geodesic equations for both  massless and massive objects in this self-dual spacetime. Using these equations we then derive in details  the effects of the polymeric function $P$  to  observations conducted in the Solar System, including the deflection angle of light by the Sun, gravitational time delay, and perihelion advance. The upper bounds on $P$ are obtained by comparing the theoretical predictions with  observational data. Then, in Sec. IV,  we study a spinning object in the self-dual spacetime and derive the geodetic procession of its spin vector, from which we obtain the constraints on the polymeric function $P$ by using the Gravity Probe B and lunar laser ranging data.  A brief summary of our main results and some discussions are presented in Sec. V.

\section{Equation of motion  for test particles in the self-dual spacetime \label{secrot}}
\renewcommand{\theequation}{2.\arabic{equation}} \setcounter{equation}{0}
	
We start with a brief introduction of the effective self-dual spacetime, which arises from the quantization of a symmetry reduced spacetime in LQG. The metric of the self-dual spacetime is given by \cite{LQG_BH}
	\begin{equation}\label{1}
	ds^2= - f(r)dt^2 + \frac{dr^2}{g(r)} + h(r)(d\theta^2+\sin^2\theta d\phi^2),
	\end{equation}
	where the metric functions $f(r)$, $g(r)$, and $h(r)$ are given by
	\begin{eqnarray}
	f(r)&=&\frac{(r-r_+)(r-r_-)(r+r_*)^2}{r^4+a_0^2},\nonumber\\
	g(r)&=&\frac{(r-r_+)(r-r_-)r^4}{(r+r_*)^2(r^4+a_0^2)},\nonumber\\
	h(r)&=&r^2+\frac{a_0^2}{r^2}.
	\end{eqnarray}
Here $r_+=2 M/(1+P)^2$ and $r_{-} = 2 M P^2/(1+P)^2$ denote the locations of  the two horizons, and $r_{*}\equiv \sqrt{r_+ r_-} = 2 MP/(1+P)^2$ with $M$ denoting the ADM mass of the solution,
 and $P$ being the polymeric function
\bqn
P \equiv \frac{\sqrt{1+\epsilon^2}-1}{\sqrt{1+\epsilon^2}+1},
\eqn
where $\epsilon$ denotes a product of the Immirzi parameter $\gamma$ and the polymeric parameter $\delta$, i.e., $\epsilon=\gamma \delta \ll 1$. From the considerations of black hole entropy \cite{KM04}, the Immirzi parameter is
 determined to be $\gamma \simeq 0.2375$. The parameter 
\bqn
a_0 = \frac{A_{\rm min}}{8\pi},
\eqn
\red{where $A_{\rm min}$} is the minimum area gap of LQG.   \red{Here we would like to emphasize that there exists a lot of choices of the value of $\gamma$ from different considerations, see \cite{Achour:2014rja, Frodden:2012dq, Achour:2014eqa, Han:2014xna, Carlip:2014bfa, Taveras:2008yf} and references therein. For example, its value can even be complex \cite{Achour:2014rja, Frodden:2012dq, Achour:2014eqa, Han:2014xna, Carlip:2014bfa} or considered as a scalar field which value would be fixed by the dynamics \cite{Taveras:2008yf}. In this paper, in order to derive the observational constraints on the polymeric parameter $\delta$ derive from the constraints on $P$, we adopt the commonly used value $\gamma = 0.2375$  from the black hole entropy calculation \cite{KM04}. Thus it is important to mention that the constraints on $\delta$ we obtained in the following sections should depend on the choice of the value of $\gamma$ .} 

By taking $a_0=0=P$, it is easy to see that the above solution reduces to the Schwarzschild black hole exactly. According to \cite{Modesto:2009ve,Sahu:2015dea}, it is natural to assume that the minimal area \red{gap} in LQG is $A_{\min} \simeq 4 \pi \gamma \sqrt{3} l_{\rm Pl}$ with $l_{\rm Pl}$ being the Planck length.  \red{In this sense, $a_0$ is proportional to $l_{\rm Pl}$ and thus is expected to be negligible. On the other hand, in order to explore the effects of $a_0$ and $P$ in the solar system, it is natural to expand (\ref{1}) in power of $1/r$. It is clearly to see that the maximal corrections from parameter $P$ is at the order of $1/r$ while $a_0$ is at $1/r^4$.} Thus, phenomenologically,  the effects of $a_0$  are expected to be very small at the scale of the Solar System, so we can safely set $a_0=0$.  
	
\section{Classical tests of the self-dual spacetime}
\renewcommand{\theequation}{3.\arabic{equation}} \setcounter{equation}{0}

Let us first consider the evolution of a massive particle in the self-dual spacetime. We start with the Lagrangian of the particle,
\bqn
\mathcal{L} = \frac{1}{2}g_{\mu \nu} \frac{d x^\mu} {d \lambda } \frac{d x^\nu}{d \lambda},
\eqn
where $\lambda$ denotes the affine parameter of the world line of the particle. For massless particles we have $\mathcal{L}=0$ and for massive ones we have $\mathcal{L} <0$. Then the geodesic motion of a particle is governed by the Euler-Lagrange equation, 
\bqn
\frac{d}{d\lambda} \left(\frac{\partial \mathcal{L}}{\partial \dot x^\mu}\right) - \frac{\partial \mathcal{L}}{\partial x^\mu}=0,
\eqn
 where a dot denotes the derivative with respect to the affine parameter $\lambda$. Then the the generalized momentum $p_\mu$ of the particle can be obtained via
\bqn
p_{\mu} = \frac{\partial L}{\partial \dot x^{\mu}} = g_{\mu\nu} \dot x^\nu,
\eqn
which leads to four equations of motions for a particle with energy $\tilde E$ and angular momentum $\tilde l$,
\bqn
p_t &=& g_{tt} \dot t  = - \tilde{E},\\
p_\phi &=& g_{\phi \phi} \dot \phi = \tilde{l}, \\
p_r &=& g_{rr} \dot r,\\
p_\theta &=& g_{\theta \theta} \dot \theta.
\eqn
From these expressions we obtain 
\bqn
\dot t = - \frac{ \tilde{E}  }{ g_{tt} } = \frac{\tilde{E}}{f(r)},\\
\dot \phi = \frac{  \tilde{l}}{g_{\phi\phi}} = \frac{\tilde{l}}{h(r) \sin^2\theta}.
\eqn
Note that one has $ g_{\mu \nu} \dot x^\mu \dot x^\nu = \varepsilon$ with $\varepsilon=-1$ for timelike geodesics and $\varepsilon= 0$ for null geodesics. Then, we find
\bqn
g_{rr} \dot r^2 + g_{\theta \theta} \dot \theta^2 &=& \varepsilon - g_{tt} \dot t^2  - g_{\phi\phi} \dot \phi^2\nb\\
&=& \varepsilon+\frac{\tilde{E}^2}{f(r)}- \frac{\tilde{l}^2}{h(r)}.
\eqn

Since we are mainly interested in the evolution of the particle in the equatorial circular orbits,  we  will set  $\theta=\pi/2$ and $\dot \theta=0$. Then the above expression can be simplified into the form
\bqn
\dot r ^2 = \tilde E^2 - V_{\rm eff} ,
\eqn
where $V_{\rm eff}$ is the effective potential of the particle, which is defined as
\bqn
V_{\rm eff} = \tilde E^2 - \left( \varepsilon+\frac{\tilde{E}^2}{f(r)}- \frac{\tilde{l}^2}{h(r)}\right) g(r).\lb{Veff}
\eqn
Then by using $\dot \phi = \tilde l/h(r)$, one obtains
\bqn\lb{rphi}
\left(\frac{dr}{d\phi}\right)^2 =\left[ \frac{h^2(r)}{\tilde l^2} \varepsilon + \frac{\tilde E^2 h^2(r)}{\tilde l^2 f(r)} - h(r)\right] g(r).
\eqn
In the following, we shall apply  this equation to the calculations of the light deflection angle, gravitational time delay, and perihelion advance in the self-dual spacetime.

\subsection{Light deflection angle}

Let us first investigate the light deflection angle in the self-dual spacetime. We start from Eq.~(\ref{rphi}), in which we have $\varepsilon=0$ for light. Introducing the impact parameter
\bqn
b\equiv \frac{\tilde l}{\tilde E}, 
\eqn
we find  Eq.~(\ref{rphi}) reduces to 
\bqn\lb{phir}
\frac{d\phi}{dr} = \pm \frac{1}{\sqrt{h(r)g(r)}} \left[\frac{h(r)}{b^2 f(r)} -1\right]^{-1/2},
\eqn
where $\pm$ correspond to increasing and decreasing $r$, respectively. Then, the distance of the closest path $r_0$ is defined as $dr/d\phi|_{r=r_0}=0$, for which we have
\bqn
b^2 = \frac{h(r_0)}{f(r_0)}.
\eqn
The light trajectory is deflected by an angle, 
\bqn
\Delta \phi &=& 2 \int_{r_0}^{+\infty} \frac{d\phi}{dr}dr - \pi,
\eqn
with $d\phi/dr$ being given by (\ref{phir}). Considering the weak field approximation and then expanding the above integral in terms of the polymeric function $P$, one obtains the deflection angle of the light,
\bqn
\Delta\phi &\simeq &\frac{4 M}{r_0} \Big(1-2 P+\mathcal{O}(P^2)\Big)\nb\\
&=&\Delta \phi_{\rm GR} (1-2 P).
\eqn

To obtain the experimental constraints from the light deflection experiment by the Sun, let us expression the deflection angle $\Delta \phi$ in terms of values of $\Delta \phi^{\rm GR}$ for the Sun,
\bqn
\Delta \phi = 1.75'' (1-2 P).
\eqn
The best available measurement of the  solar gravitational deflection comes from the astrometric observations of quasars on the solar background performed with the very-long baseline interferometry (VLBI) \cite{VLBI_deflection}, 
which leads to the constraint on the polymeric function $P$,
\bqn
-2.5 \times 10^{-5} < P< 1.25 \times 10^{-4}  \;\;\;\; (68\% \;\;{\rm C.L.}). ~~~~~
\eqn
Considering $P>0$, thus one has
\bqn
0< P < 1.25 \times 10^{-4} \;\;\;\; (68\% \;\;{\rm C.L.}).
\eqn
For $\gamma = 0.2375$, the above constraint can be transformed to a constraint on the polymeric parameter $\delta$ as
\bqn
| \delta| <  0.0942 \;\; \;\;(68\% \;\;{\rm C.L.}).
\eqn
\red{It is worth noting that the above constraint is consistent with that obtained in \cite{Sahu:2015dea} using the VLBI data in \cite{add_VLBI}. }

\subsection{Gravitational Time Delay}

We consider   the time delay where a radar signal is sent from Earth or spacecraft pass to the Sun and reflect off another planet or spacecraft. The time delay can also be studied by using Eq.(\ref{phir}), from which one obtains
\bqn
\frac{dt}{dr} &=& \frac{dt}{d\phi} \frac{d\phi}{dr} = \frac{d\phi}{dr} \frac{\dot t}{\dot \phi}\nb\\
&=& \pm  \frac{1}{ b } \frac{1}{\sqrt{f(r)g(r)}} \left[\frac{1}{b^2} -\frac{f(r)}{h(r)}\right]^{-1/2}.
\eqn
Then the time spent by a radar signal that travels from the Sun to the point $r_A$ can be obtained by performing the integral
\bqn
 t(r_A) =  \frac{1}{ b } \int_{r_0}^{r_A}  \frac{1}{\sqrt{f(r)g(r)}} \left[\frac{1}{b^2} -\frac{f(r)}{h(r)}\right]^{-1/2} dr. ~~~~~
\eqn
Again considering the weak field approximations,  one  finds
\bqn
 t(r_A) &\simeq & \sqrt{r_A ^2- r_0^2} + M \sqrt{\frac{r_A -r_0}{r_A+r_0}} +2 M {\rm arccosh}\left(\frac{r_A}{r_0}\right) \nb\\
 && - 4 M P \left( \sqrt{\frac{r_A -r_0}{r_A+r_0}} + {\rm arccosh}\left(\frac{r_A}{r_0}\right)\right).
\eqn

Then the time delay of a radar signal that is sent from Earth or spacecraft and then reflects off another planet or spacecraft can be divided into two cases, the inferior conjunction and superior conjunction. In the inferior conjunction case,  the planet (or spacecraft, denoted by B), which reflects the radar signal, is located between the Earth (or spacecraft, denoted by A) and the Sun. For this case,  the time delay due to the self-dual spacetime can be obtained by
\bqn
\Delta t_I \simeq 4 M \ln\frac{r_A}{r_B}  \times (1- 2 P) = \Delta t_I^{\rm GR} (1- 2P).\nb\\
\eqn
In the superior conjunction case, the planet   that reflects the radar signal and the Earth   is on opposite sides of the Sun, and the time delay for this superior conjunction case can be written as
\bqn
\Delta t_{S} &\simeq& 4 M+4 M \ln \frac{4 r_A r_B}{r_0^2} - 16 MP - 8 MP \ln \frac{4 r_A r_B}{r_0^2} \nb\\
&=& \Delta t_{S}^{\rm GR}  - 16 MP - 8 MP \ln \frac{4 r_A r_B}{r_0^2}.
\eqn
Here we use the experimental results of the Cassini satellite for the time delay to constrain the polymeric function in the self-dual spacetime \cite{cassini}. The Cassini experiment does not measure the time delay 
directly, but instead the relative change in the frequency in the superior conjunction case,
\bqn
\delta \nu = \frac{\nu(t) - \nu_0}{\nu_0} = \frac{d}{dt} \Delta t_{S},
\eqn
where $\nu_0$ is the frequency of the radio waves emitted from the Earth and then t being  reflected back to the Earth at the frequency $\nu(t)$. Hence,  the relative shift in the frequency is given by
\bqn
\delta \nu  \simeq - \frac{8 M (1- 2 P) }{r_0} \frac{dr_0(t)}{dt}.
\eqn
The Cassini experiment measures the frequency shift for approximately 25 days, where 12 days before and 12 days after the superior conjunction. During one day the distance of the closet approach of the radio waves changes by about $1.5 R_{\odot}$, where $R_{\odot}$ denotes the radius of the Sun. Thus,  the frequency shift induced by the polymeric function $P$ is
\bqn
\delta \nu_P \simeq \frac{256}{27} P \frac{M_{\odot}}{R_{\odot}} v_{E},
\eqn
in which $v_E = dr_0/dt$ is the velocity of the Earth. In the Cassini experiment, the accuracy of the relative shift in the frequency is $10^{-14}$ \cite{cassini}, from which one obtains the constraint 
\bqn
\delta \nu_P < 10^{-14},
\eqn
which leads to
\bqn
0<P< 5.5\times 10^{-6}.
\eqn
This constraint is stronger than that obtained by the observations of the deflection angle. Similarly, if one takes $\gamma=0.2375$, the above constraint leads to the constraint to the polymeric parameter 
\bqn
|\delta | < 0.0199.
\eqn

\subsection{Perihelion Advance}

Now let us turn to the massive particles moving in the self-dual spacetime and study the perihelion advance of their orbits. We start from Eq. (\ref{rphi}) with $\varepsilon=-1$ in terms of a new variable $x=1/r$, 
which yields, 
\bqn
\left(\frac{d x}{d\phi} \right)^2= x^4 \left[ - \frac{h^2(r)}{\tilde l^2} + \frac{\tilde E^2 h^2(r)}{\tilde l^2 f(r)} - h(r)\right] g(r).\nb\\
\eqn
Differentiating it with respect to $\phi$ and then expanding the equation by assuming that $P$ is a small parameter, one finds the orbits of the massive particles are governed by the following differential equation,
\bqn
&&\frac{d^2x}{d\phi^2} + x -  \frac{M}{\tilde l^2} \simeq   3 M x^2 \nb\\
&&~~~~~~  - 4 M \left(\frac{\tilde E^2}{\tilde l^2} + \frac{2 M}{\tilde l^2} x - 4 M x^3\right) P. \lb{EE}
\eqn
The right-hand side of the above equation can be treated as perturbations to the Newtonian gravity. By ignoring the perturbation terms, the unperturbed solution of the above equation is given by
\bqn
x_0 = \frac{M}{\tilde l^2} (1+ e \cos\phi),
\eqn
which describes an elliptical orbit with the eccentricity $e$. When the perturbations in the right-hand side of (\ref{EE}) is included, the elliptical orbit acquires a small correction, i..e, $x=x_0 + x_1$, where $x_1$ satisfies 
\bqn
&& \frac{d^2 x_1}{d\phi^2} + x_1 \simeq 3M x_0^2\nb\\
&&~~~ - 4 M \left(\frac{\tilde E^2}{\tilde l^2} + \frac{2 M}{\tilde l^2} x_0 - 4 M x_0^3\right) P.
\eqn
 Substituting the solution of $x_0=M (1+e \cos\phi)/\tilde l^2$ into the above equation, one finds
 \bqn
 \frac{d^2 x_1}{d\phi^2} + x_1 = A_0 +  A_1 \cos\phi +  A_2 \cos^2 \phi +  A_3 \cos^3 \phi ,\nb\\
 \eqn
 where
 \bqn
 A_0 &=&  \frac{3 M^3}{\tilde l^4} - 4 M \left(\frac{E^2}{\tilde l^2} + \frac{2 M^2}{\tilde l^4} + \frac{4 M^4}{\tilde l^6}\right) P, \\
 A_1 &=& \frac{3 M^3}{\tilde l^4} \left(2e - \frac{8}{3} e P + 12 e P \frac{M^2}{ \tilde l^2}\right), \\
 A_2 &=&  \frac{3 M^3}{\tilde l^4}  \left( e^2 + 12 e^2 P \frac{M^2}{\tilde l^2} \right), \\
 A_3 &=& \frac{3 M^3}{\tilde l^4}  \times \frac{16}{3}e^3 \frac{M^2}{\tilde l^2}.
 \eqn
 Then the solution of $x_1$ is given by
 \bqn
 x_1 &=& A_0+\frac{A_2}{2} - \frac{A_2}{6} \cos (2 \phi) - \frac{1}{32} \cos (3 \phi)  \nb\\
 && + \left(\frac{1}{2} A_2 + \frac{3}{8} A_3\right) \phi \sin\phi.
 \eqn
 In this solution, only the last term (in the second line) contributes to the perihelion advance, thus one can ignore the other terms in the solution and finally  has
 \bqn
 x &\simeq & \frac{M}{\tilde l^2} (1+ e \cos \phi) + \left(\frac{1}{2} A_1 + \frac{3}{8} A_3\right) \phi \sin\phi \nb\\
 &\simeq &  \frac{M}{\tilde l^2} \left[1 + e \cos \left(\phi - \frac{\delta \phi_0}{2 \pi}  \phi \right)  \right] ,\lb{orbit}
 \eqn
 where
 \bqn
 \delta \phi_0 \simeq  \frac{6 \pi M^2}{\tilde l^2} \left(1- \frac{4}{3}P\right), \lb{delta_phi}
 \eqn
which is the angular shift of the perihelia per orbit. 
 
 Now we would like to eliminate the angular momentum $\tilde{l}$ from (\ref{delta_phi}). Considering the orbit along (\ref{orbit}), we find that the minimum value of $r_-$ and 
 the maximum one of $r_+$  can be obtained from (\ref{orbit}) at $(1-\delta\phi_0/2\pi)\phi=0$ and $(1-\delta\phi_0/2\pi)\phi=\pi$, respectively. Then, we find,
 \bqn
 r_-= \frac{\tilde l}{M (1+e)}, \;\;\\
 r_+=\frac{\tilde l}{M (1-e)}.
 \eqn
 Thus,  the semi-major axis $a_0$ of the ellipse is 
 \bqn
 a_0=\frac{r_-+r_+}{2} = \frac{\tilde l^2}{M (1-e^2)}. 
 \eqn
 Using this expression, the perihelion advance per orbit can be expressed as
 \bqn
 \Delta \phi = \Delta \phi^{\rm GR} \left(1- \frac{4}{3}P\right),
 \eqn
 where
 \bqn
 \Delta \phi^{\rm GR} = \frac{6\pi M}{a_0(1-e^2) }.
 \eqn
 Note that by taking $P=0$, one recovers the classical result for the Schwarzschild spacetime. 
 
 Let us now consider observational constraints that can be imposed on the polymeric parameter $P$. We first consider the observation of the anomalous perihelion advance for Mercury. The current most accurate detection was done by the MESSENGER mission \cite{message}, in which the contribution from the Schwarzschild-like procession is measured to be
 \bqn
 \Delta \phi = (42.9799 \pm 0.0009)'' /{\rm century}.
 \eqn
 We use the observational error in experimental data to compute upper-bounds for the polymeric parameter $P$. For the motion of Mercury around the Sun, the observational error is $0.009''/{\rm century}$. One expects  that the contribution from LQG is less than the observational error. This procedure leads to a bound on the polymeric function $P$ as
 \bqn
 0<P<1.57\times 10^{-5}.
 \eqn
 From this bound, the polymeric parameter $\delta$ is constrained to be
 \bqn
 |\delta| < 0.033.
 \eqn
\red{In the above calculation, we have ignored the contributions to to the perihelion advance from the Lense-Thirring effects due to the angular momentum of the Sun. This effect is proportional to the angular momentum of the Sun and is given by \cite{LT}
 \bqn
 \Delta \phi^{\rm LT} = - \frac{6 S_{\odot}\cos i}{a_0^3 (1-e^2)^{3/2}},
 \eqn
 where $S_{\odot}$ is the angular momentum of the Sun and $i$ is the inclination of the solar equator to Mercury’s orbit plane. According to the analysis in \cite{message}, the contribution of $\Delta \phi^{\rm LT}$ per century is smaller than the uncertainty in the measurement of the perihelion advance per century, thus for the purpose of constraining LQG effects in this work we ignore the effects of $\Delta \phi^{\rm LT}$ and only consider the static case. }

We then turn to consider the measured perihelion advance of LAGEOS satellites around the Earth. Using 13 years of tracking data of the LAGEOS satellites, the precession of the periapsis of the LAGEOS II satellite was measured to be \cite{LAGEOS}
\bqn
\Delta \phi = \Delta \phi^{\rm GR} \Big[1+ (0.28 \pm 2.14)\times 10^{-3}\Big],
\eqn
which corresponds to a bound on the polymeric function $P$ and parameter $\delta$ of
\bq
0<P<0.0014,
\eq
and 
\bq
|\delta| < 0.32,
\eq
respectively.

On the other hand, the observations of the stars orbiting the central black hole of the Milky Way galaxy provide a different environment to test gravity in the strong gravity regime. These stars has been observed for 27 years and now their orbital parameters can be determined very accurately. Recently, the GRAVITY collaboration has detected the Schwarzschild precession of the S2 star to be \cite{S2}
\bqn
\Delta\phi =\Delta \phi^{\rm GR}(1.1 \pm 0.19),
\eqn
where
\bqn
\Delta \phi^{\rm GR} = 12'
\eqn
per orbit period from the prediction of GR. For the LQG corrections to the procession, this detection implies
\bqn
0<P<0.0675,\;\;\; |\delta|<2.3.
\eqn

\section{Geodesic precession of spinning objects in the self-dual spacetime}
\renewcommand{\theequation}{4.\arabic{equation}} \setcounter{equation}{0}

Now let us turn to consider the evolution of a spinning particle with its four-velocity vector $u^\mu=dx^\mu/d\lambda$ and four-spin vector $s^\mu$ in the self-dual spacetime. The equation of motions of this type of particles is governed by two equations, namely, the geodesic equation
\bqn
\frac{du^\mu}{d\lambda} + \Gamma^\mu_{\nu \lambda} u^\nu u^\lambda=0,
\eqn
and the parallel transport equation
\bqn
\frac{ds^\mu}{d\lambda} + \Gamma^\mu_{\nu \lambda} s^\nu u^\lambda=0,\lb{spin}
\eqn
where the four-velocity vector $u^\mu$ and four-spin vector $s^\nu$ satisfy the orthogonal condition
\bqn
u^\mu s_\mu=0.
\eqn	
The spin vector $s^\mu$ also satisfies the normalization condition
\bqn
s^\mu s_\mu =1.
\eqn 

 Since the self-dual spacetime we considered here is a spherically symmetric spacetime, we can comfortably choose to work on the equatorial plane, i.e., with $\theta=\pi/2$ without loss of any generality. To simplify the problem, we further assume that the test spinning particle moves in a circular orbit, i..e, $\dot r = 0=\dot \theta$. Then the four velocity $u^\mu=\dot x^\mu$ can be expressed as follows in terms of the constants of motion $\tilde E$ and $\tilde l$,
 \bqn
 u^t = \dot t =\frac{\tilde E}{f(r)}, \lb{ut}\\
 u^\phi= \dot \phi = \frac{\tilde l}{h(r)}.\lb{uphi}
 \eqn
One can define the angular velocity of the spinning particle as
\bqn
\Omega = \frac{u^\phi}{u^t} = \frac{\tilde l}{\tilde E} \frac{f(r)}{h(r)}.
\eqn
Note that the radial and $\theta$ components of the four velocity vanish since $\dot r=0=\dot \theta$.  For the stable circular orbit in the equatorial plane, the effective potential $V_{\rm eff}(r)$ in (\ref{Veff}) must obey 
\bqn
\tilde E^2-V_{\rm eff}=0,\;\;\;\; \frac{dV_{\rm eff}}{dr}=0.
\eqn
Solving this two equations one obtains
\bqn
\tilde E &=& \sqrt{\frac{f^2(r)h'(r)}{f(r)h'(r) - h(r)f'(r)}},\\
\tilde l &=&  \sqrt{\frac{h^2(r)f'(r)}{f(r)h'(r) - h(r)f'(r)}},\\
\Omega &=& \sqrt{\frac{f'(r)}{h'(r)}}.
\eqn
Plugging these results into (\ref{ut}) and (\ref{uphi}) we can obtain $u^t$ and $u^\phi$ of the test spinning particle in equatorial circular orbits. Then in the self-dual spacetime, the parallel transport equation (\ref{spin}) along the circular orbits with radius $r$ in the equatorial plane reads
\bqn
&& \frac{ds^t}{d\lambda} +  \frac{1}{2}  \frac{f'(r)}{f(r)} u^t s^r =0,  \lb{st} \\
&& \frac{ds^r}{d\lambda} +\frac{1}{2} g(r) f'(r) u^t s^t - \frac{1}{2} g(r) h'(r) u^\phi s^\phi=0, ~~~~~~ \lb{sr}\\
&&\frac{ds^\theta}{d\lambda} =0, \\
&&\frac{ds^\phi}{d\lambda} +\frac{1}{2} \frac{h'(r)}{h(r)} u^\phi s^r=0.\lb{sphi}
\eqn
Differentiating  (\ref{sr}) with respect to the affine parameter $\lambda$ and converting $\lambda \to t$ using the relation $d t  = u^t d \lambda$, one arrives at a second-order ordinary differential equation of $s^r$,
\bqn
\frac{d^2s^r}{d t^2} + \frac{1}{4}\left[\frac{g(r)h'^2(r)}{h(r)} \Omega^2- \frac{g(r)f'^2(r)}{f(r)}  \right] s^r=0,\nb\\ \lb{ss}
\eqn
which can be solved to yield,
\bqn
s^r(t) = s^r(0) \cos (\omega_g t),
\eqn
where 
\bqn
\omega_g= \frac{1}{2} \sqrt{\frac{g(r)h'^2(r)}{h(r)} \Omega^2- \frac{g(r)f'^2(r)}{f(r)}} \lb{omega_g},
\eqn
is the frequency of the oscillation pertaining to the spin four-vector $s^\mu$. Note that in deriving (\ref{ss}) we have used (\ref{st}) and (\ref{sphi}). Given this solution for the radial component $s^r$ and one can immediately solve for $s^t$, $s^\theta$, and $s^\phi$, yielding
\bqn
s^t( t) &=& -\frac{1}{2}\frac{f'(r)}{f(r)} s^r(0) \sin (\omega_g t), \\
s^\theta(t) &=&0,\\
s^\phi(t) &=& -\frac{1}{2} \frac{h'(r)}{h(r)} \Omega s^r(0) \sin(\omega_g t).
\eqn
Here we have imposed the initial conditions such that the spin vector was initially directed along the radial direction, i.e.,  $s^t(0) =s^\phi(0)=s^\theta(0) =0$. 

By the inspection of the expression (\ref{omega_g}), it is evident that the the angular velocity $\omega_g$ of rotation of the spin vector is different from the angular velocity of the massive spinning particle along the circular orbit. It is this difference that leads to a procession of the spin vector. To see this clearly, let us compare $\omega_g$ and $\Omega$ by expanding (\ref{omega_g}) in terms of $M$ and $P$ as
\bqn
\frac{\omega_g}{\Omega} &=& \frac{1}{2} \sqrt{\frac{g(r)h'^2(r)}{h(r)} - \frac{g(r)f'(r) h'[r]}{f(r)}} \nb\\
&\simeq& 1-\frac{3 M}{2r} + \frac{2 M}{r} P,
\eqn
which shows clearly $\omega_g < \Omega$. This implies that when the spinning particle completes one rotation along the circular orbit, the spin vector has not yet completed a complete circle. This phenomenon is called {\em geodetic procession}. For one complete period of the circular orbit, the angle of the geodetic procession can be expressed as
\bqn
\Delta \Theta &=& 2 \pi \left(1- \frac{\omega_g}{\Omega} \right) \nb\\
&\simeq &\frac{ 3 \pi M}{r} \left(1- \frac{4}{3} P \right ),
\eqn
where the second term in the bracket represents the corrections from the LQG effects in the self-dual spacetime. It is transparent that the geodetic precession angle $\Delta \Theta$ decreases with the polymeric function $P$. When $P=0$ the above geodetic precession angle $\Delta \Theta$ reduces to the result for the Schwarzschild spacetime. 

The geodetic procession can be tested by using gyroscopes in the near-earth artificial satellites, which has been detected by the Gravity Probe B \cite{GPB}.  Considering that the Gravity Probe B was spaced at an attitude of 642 km and had an orbital time period of 97.65 min, the geodetic effect leads to a procession of the gyroscope  spin axis by 6,606.1 milliarcseconds (mas) per year, as predicted by GR.  This procession is measured by the Gravity Probe B to be \cite{GPB}
\bqn
\Delta \Theta = (6601.8 \pm 18.3 ) {\rm mas}/{\rm year}.
\eqn
This measurement leads to a bound on the polymeric function $P$ of
\bqn
0<P<2.6\times 10^{-3},
\eqn
which corresponds to a bound on the polymeric parameter $\delta$ of
\bqn
|\delta| < 0.43.
\eqn

The Earth-Moon system in the field of the Sun can also be considered as a gyroscope. This makes it is possible to detect the geodetic procession by measuring the Lunar orbit by using the Lunar laser ranging data. Recent
 measurement of the geodetic procession yields a relative deviation from GR as \cite{lunar}
\bqn
\frac{\Delta \Theta - \Delta \Theta^{\rm GR}}{ \Delta \Theta^{\rm GR}} = -0.0019 \pm 0.0064.
\eqn 
From this result one can get the bound of the polymeric function $P$ of
\bqn
0<P< 6.2\times 10^{-3},
\eqn
which corresponds to the bound on the polymeric parameter $\delta$ of
\bqn
|\delta| < 0.67.
\eqn

\section{Summary and Discussions}
\renewcommand{\theequation}{5.\arabic{equation}} \setcounter{equation}{0}

\begin{table*}
\caption{Summary of estimates for upper bounds of the polymeric function $P$ and the parameter $\delta$ in the self-dual spacetime from several observations. }
\lb{table}
\begin{ruledtabular}
\begin{tabular} {cccc}
 Experiments/ Observations & $P$& $|\delta|$   & Datasets \\
 \hline
Light deflection & $1.25\times 10^{-4}$ & 0.0942 & VLBI observation of quasars \cite{VLBI_deflection} \\
Time delay & $ 5.5\times 10^{-6}$ &0.0199 & Cassini experiment \cite{cassini}  \\
Perihelion advance & $1.57\times 10^{-5}$& 0.033& MESSENGER mission \cite{message} \\
\; & $1.4 \times 10^{-3}$ &0.32 & LAGEOS satellites \cite{LAGEOS} \\
\; & $0.0675$ &2.3 & Observation of S2 star at Galactic center \cite{S2} \\
Geodetic procession & $2.6\times 10^{-3}$& 0.43& Gravity Probe B \cite{GPB} \\
\; & $6.2 \times 10^{-3}$& 0.67&  Lunar laser ranging data \cite{lunar} \\
\end{tabular}
\end{ruledtabular}
\end{table*}

LQG provides an elegant resolution of both the classical big bang and black hole singularities. Recently, a regular static spacetime, the {\em self-dual spacetime}, is derived from the mini-superspace approach, based on the polymerization quantization procedure in LQG \cite{LQG_BH}. In this paper, we study the observational constraints that can be imposed on the polymeric function $P$ arising from LQG. For this purpose, we calculate theoretically the  effects of the polymeric function $P$ to some astronomical observations conducted in the Solar System, including the deflection angle of light by the Sun, gravitational time delay, perihelion advance, and geodetic procession. Confronting the theoretical predictions with the observations, we derive the upper bound on the polymeric function in the self-dual spacetime. Our results are summarized in Table.~\ref{table}.

It is remarkable that the measurement of the gravitational time delay by the Cassini experiment provides by far the most sensitive tool to constrain the effects of LQG in the Solar System. This measurement gives the tightest constraints [cf. in Table.~\ref{table}] on the polymeric function $P$ of $0<P<5.5\times 10^{-6}$ and on polymeric parameter $\delta$ of $|\delta| <0.0199$. Another important constraint comes from the observation of the perihelion advance for Mercury by the MESSENGER mission, which leads to an upper bound on $P$ of $1.57 \times 10^{-5}$. In the near future, the accuracy of the measurement for the Mercury's perihelion advance will be significantly improved by the joint European-Japanese BepiColombo project, which was launched in October, 2018. It is expected that this mission will improve the accuracy of the perihelion advance to be $10^{-4} \;{\rm as /century}$ \cite{will2018, Bepi}, which is one order of magnitudes better than the current accuracy of about $10^{-3} {\rm as /century}$ \cite{message}. With this mission, one can improve the constraints on the polymeric function $P$ to $0<P \lesssim 2 \times 10^{-6}$, which is much more restricted  than that obtained from the Cassini experiment. 

We also calculate the effects of the polymeric function on the geodetic procession of a spinning object in the self-dual spacetime. The observation constraints on $P$ has also been derived from the Gravity Probe B data and the Lunar laser ranging data. Although these constraints are not as tighter as those obtained from  the observations of the light deflection angle, gravitational time delay, and perihelion advance of Mercury, they do provide a different and interesting window  
 to explore the features of the self-dual spacetime.

\section*{Acknowledgements}

T.Z. is supported in part by National Natural Science Foundation of China with the Grants No.11675143, the Zhejiang Provincial Natural Science Foundation of China under Grant Nos. LR21A050001, LY20A050002, and the Fundamental Research Funds for the Provincial Universities of Zhejiang in China under Grants No. RF-A2019015.  A.W. is supported by National Natural Science Foundation of China with the Grants Nos. 11675145 and 11975203.

\end{document}